# Graphene - a rather ordinary nonlinear optical material.

*J. B. Khurgin*

*Johns Hopkins University, Baltimore, MD 21218, USA*

**Abstract**

An analytical expression for the nonlinear refractive index of graphene has been derived and used to obtain the performance metrics of third order nonlinear devices using graphene as a nonlinear medium. None of the metrics is found to be superior to the existing nonlinear optical materials.

Nonlinear optics, tracing its origins to the invention of laser in 1960's [1-3] has been an exciting field in the last 50 plus years: significant advances have been made in understanding nonlinear phenomena, and a number of practical applications have emerged, such as in frequency conversion and optical parametric generation for the second order nonlinear phenomena, and mode locking, comb generation, and a variety of nonlinear spectroscopic tools for the third order nonlinear effects. Yet the most enticing promise of nonlinear optics – that of controlling light by light and hence all optical switching and computing -- remains unfulfilled because the nonlinear susceptibilities of the host of materials that have been explored to date are simply too small to support efficient all-optical switching. Efficient switching requires one to achieve either a nonlinear phase shift commensurate with 180 degrees (or absorption change by 90% or so) over a distance that is not much longer than one absorption length, and, when it comes to fast (say sub-picosecond) nonlinearities, the best results have been obtained with optical fibers, but only due to long propagation length, which makes these schemes impractical. Although all-optical switching has been demonstrated in a variety of waveguides, such as silicon or chalcogenide, the relatively large length and high power requirements prevented these materials from becoming practical.

The nonlinear index of the best nonlinear materials in the optical/near IR range is on the order of $n_2 \leq 10^{-13} cm^2/W$ [4] which follows from very simple considerations: the electrons are confined within the bonds by a potential that is parabolic near the equilibrium and becomes non-parabolic only when the externally applied field approaches the intrinsic field whose magnitude is on the scale of $E_i \sim 10^8 V/cm$ which immediately provides the right scale for $n_2 \sim 2\eta_0 E_i^{-2}$. These scaling relations are fundamental for all bound electrons; hence it would be unrealistically optimistic to expect a breakthrough in conventional materials with bound (valence) electrons.

When it comes to free electrons in either free space or condensed matter the situation is analogous. Low energy electrons are all subject to the parabolic dispersion relation between the energy and momentum, this relation becomes non-parabolic (and nonlinearity ensues) only when the energy of electron approaches some "threshold value" , e.g the Fermi energy in metals, bandgap energy in doped semiconductors [5-7] or relativistic energy $mc^2$ for free electrons [8]. As a result, the nonlinear index even in narrow bandgap materials, such as InAs is insufficient to achieve switching in less than the absorption length.

It is hardly surprising then, that each time a new class of materials becomes available, there is immediately a healthy impulse to explore the possibility that in these materials optical nonlinearities, often referred to as "giant", will exist and the old curse of "insufficient for optical switching ratio of $n_2$ to the absorption coefficient" will be finally lifted. This had been the case with polymers in 1970's [9], semiconductor quantum wells [10] and superlattices [11] in 1980's , quantum dots in 1990's [12], and, more recently, with a multitude of metal-incorporated plasmonic and meta-materials [13]. Each time though, the initial enthusiasm has waned when faced with indisputable realities of nonlinearity being controlled by a very few parameters (essentially only the bandgap, frequency, and optical transition matrix elements), and, in the end, not being sufficient for all optical switching due to the small total phase shift, excessive absorption loss, optical damage, or, most often, the combination of all three. To this day, the list of practical materials for all optical switching remain relatively short, and none of them works sufficiently well to develop low power all-optical switches with a footprint small enough for applications in integrated optics [14].

Thus, it was expected that once graphene, a new form of carbon, appeared on the stage in early 2000's [15], its nonlinear optical properties would be explored and claims of "giant" nonlinearity would be made. Indeed, one did not have wait long before a large number of reports of extraordinary large nonlinearity in graphene appeared. Huge values of nonlinear susceptibility were estimated [16-18] and even measured [18-20], and predictions of various switching schemes [21], including single photon optical switching [22] were made, but when it comes to practical optical switching, progress has been less obvious, and, other than passive mode locking of lasers [20], which does not require large change in index or absorption, there is no report of working graphene devices with performance exceeding that of more conventional materials. It is the goal of this paper to establish whether graphene is indeed a better material when it comes to ultrafast all-optical switching. It is important to stress that, for all optical switching, a large value of nonlinear index of refraction is definitely not enough. The one and only figure of merit is the maximum amount of phase shift that can be attained over a distance that is less than one absorption length. Two processes can limit this figure: first, and most obvious, the optical absorption itself, and second, saturation of the nonlinearity. And both of these processes play important roles in graphene.

Optical transitions in graphene can be both interband and (in doped graphene) intraband and both types engender nonlinearities. The first thing that can be said about the **interband nonlinearity** [17-20] is that it **has absolutely nothing to do with Dirac-like dispersion of electrons** and is no different from the optical nonlinearity of any 2D electron systems excited far above the bandgap, as in, for instance SESAM [23] saturable absorption. In fact in both [18] and [19] the nonlinearity scales linearly with the number graphene monolayers, even though Dirac dispersion no longer exists after two monolayers. The similarity between the interband transitions in graphene and any narrow gap semiconductor is evident if one simply takes a look at the dispersion in the conduction band of, say InAs away from the bandgap. It is almost linear with the slope, i.e. the electron velocity approaches $1.1\times10^8$ cm/s [24,25] comparable to (in fact larger than) graphene, as shown in Fig1.a. Clearly, the behavior of the electron with a high enough energy, say 1eV in the band cannot be affected by the presence (or absence) of the bandgap.

The nonlinear index for the interband transitions can become very large due to the resonant enhancement, but so does the absorption coefficient. In [18] the equivalent 3D nonlinear refractive index for the graphene was found to be as high as $10^{-9} cm^2/W$ but the equivalent absorption is about $7\times10^5 cm^{-1}$. The phase shift attainable over one absorption length of 15nm for the 1.5μm light is $\Delta\Phi = 2\pi n_2 I L_{abs}/\lambda \sim 0.25\times10^{-10} I$ where $I$ is the intensity in W/cm². Therefore, even with intensities as high as 1 GW/cm² $\Delta\Phi$ is almost two orders of magnitude less than the desired 180 degrees. For comparison, in chalcogenide waveguides with $n_2 \sim 10^{-13} cm^2/W$ one can achieve $\pi$-shift in a few mm long waveguide. The maximum phase shift in graphene obviously does not change if one uses waveguide geometry since both linear absorption and nonlinear susceptibility will be equally diluted. This result is easily predictable since according to Eq (4) in [18] the ratio between linear interband sheet conductivity of graphene $\sigma^{(1)} = e^2/4\hbar$ and the third order sheet conductivity is roughly $\sigma^{(3)}/\sigma^{(1)} \sim 1/3E_i^2$ where $E_i = \hbar\omega^2/ev_F \approx 10^7 V/cm$ **is the same "intrinsic field" that** is characteristic of any nonlinear process. Since nonlinear phase shift is proportional to $\eta_0\sigma^{(3)}E^2N_l$ and absorption to $\eta_0\sigma^{(1)}N_l$ (where $\eta_0 = 377\Omega$ is a vacuum impedance and $N_l$ is the number of layers) the ratio $\sigma^{(3)}/\sigma^{(1)}$ is precisely the phase shift per one absorption length, and for it to be comparable to $\pi$ the optical field must reach values of the order of $E_i$-corresponding to intensities of 100GW/cm². Hence **nominally huge values of nonlinear coefficients for interband transitions in the graphene** **are solely due to huge absorption coefficient** and in practice do not lead up to any improvement in optical switching relative to existing nonlinear optical materials. Of course, graphene may find a niche as a fast saturable absorber for mode-locking but due only to its ability to operate over wide range of frequencies, particularly long wavelengths, and not because of better performance than say SESAM [23].

Let us now turn our attention to the interband optical nonlinearity in doped graphene [16,21,22], shown schematically in Fig.1b, which unlike the interband counterpart is indeed caused by the strong non-parabolicity of the Dirac electrons. The origin can be easily understood from Fig2.a where the color corresponds to the projection of the electron velocity $v_x = v\cos\theta$ onto the x axis along which the AC electric field $E_\omega \sin\omega t$ is applied. The conduction electrons whose sheet density is $N_{2D}$ are confined within the Fermi circle of radius $k_F = \pi^{1/2} N_{2D}^{1/2}$. In the electric field the center of the Fermi circle oscillates as $k_0(t) = (eE_\omega / \hbar\omega)\cos\omega t$ -clearly when the electrons pass near the origin (Dirac point) the velocity quickly changes from negative to positive engendering a large AC current. The instant sheet current density, as a function of instant position of center of Fermi circle can be found as $\boldsymbol{J} = \pi^{-1} e v_F k_F \boldsymbol{k}_0 F_\sigma(k_0/k_F)$, where

$$F_\sigma(x) = \frac{1}{\pi x} \int_0^1 z \int_0^{2\pi} \frac{x + z\cos\varphi}{\sqrt{x^2 + z^2 + 2xz\cos\varphi}} d\varphi dz \tag{1}$$

is a normalized conductivity (since $\sigma \sim J/k_0$) function shown in Fig.2b which clearly saturates with $F_\sigma(2) \approx 1/2$, and, in fact can be very well approximated a simple saturation function $F_{\sigma 1} = 1/(1 + x^2/4)$ shown as dotted line. Also shown in Fig2c is the plot of $F_\sigma(x^2)$ which shows the shape of saturation of conductivity as a function of intensity ($I_\omega \sim k_0^2$). The sheet current density J also saturates as shown in Fig.2d for two values of current density. With higher carrier density current is larger and it saturates at larger values of $k_0$, i.e. at higher field. This can be also seen from Fig2e where the sheet conductivity normalized to the frequency $J/k_0 = (\hbar\omega/e)\sigma_\omega = \pi^{-1}ev_F k_F F_\sigma(k_0/k_F)$ is plotted. Clearly, at low doping densities the conductivity is highly nonlinear –which is easy to understand – the conductivity changes rapidly only in the vicinity of the Dirac point, and at low doping densities most electrons are near the Dirac point and thus behave nonlinearly. But overall conductivity is small. For higher doping densities the conductivity is higher, but it does not saturate quickly and remains linear even at high fields, with saturation setting in at $k_0 \approx 2k_F$ (dots shown in Fig.2d,e), or at saturation field strength $E_{sat,\omega} \sim 2k_F(\hbar\omega/e) = 2\omega E_F/ev_F$, which plays essentially the same role as intrinsic field for interband transitions. Herein, as we shall soon see, lies the conundrum of graphene as nonlinear material: electrons near the Dirac point are extremely nonlinear, but there are simply too few of them to provide a meaningful nonlinear effect, such as a phase shift sufficient for switching.

To demonstrate, we first introduce the linear 2D susceptibility of the graphene sheet as

$$\chi_\omega^{(1)} = -j\sigma_\omega(0)/\omega\varepsilon_0 = -\frac{1}{\pi}\frac{e^2 E_F}{\varepsilon_0 \hbar^2 \omega^2} \tag{2}$$

Then the entire field dependent susceptibility can be written as

$$\chi_\omega(I_\omega) = \chi_\omega^{(1)} F_\sigma(E_\omega/E_{sat,\omega}) = \chi_\omega^{(1)}/(1+I_\omega/I_{sat,\omega}) \tag{3}$$

where the saturation intensity is $I_{sat,\omega} = E_{sat,\omega}^2/2\eta_0 \sim 2(\omega E_F/ev_F)^2/\eta_0$. One can now estimate the effective 3D index as $n_{eff,\omega} = \chi_\omega/2d_{eff}$, where $d_{eff}$ is the effective thickness. For light propagating in the direction normal to the layers one typically uses $d_{eff} \approx 0.33 nm$, while for light in the waveguide containing $N$ graphene monolayers $d_{eff} \approx d_g/N$ where $d_g$ is the effective thickness of the waveguide. Then for small values of intensity one can expand (3) to obtain $n_{eff,\omega}(I_\omega) = n_{eff,\omega} + n_{2,eff} I_\omega$, where the linear effective index is $n_{eff,\omega} = -2\alpha_0(E_F/\hbar\omega)(c/\omega d_{eff})$, where $\alpha_0$ is a fine structure constant, while the nonlinear index is $n_{2,eff} = 4\pi\alpha_0^2 c v_F^2 / E_F \omega^4 d_{eff}$.

Two caveats should be mentioned: first of all, to avoid excessive interband absorption the condition $\hbar\omega < 2E_F$ applies (even though in reality, especially at room temperature, optical phonon assisted mid-gap absorption [26] will limit the range of frequencies for well below $2E_F$). Also, at low frequencies the free carrier scattering will set in, effectively placing a lower boundary on the frequency range, and, in all our expression one should use $\omega^2 + \tau^{-2}$ in place of $\omega^2$, where $\tau \sim 100 fs$ is the scattering time. The results for the nonlinear refractive index are shown in Fig. 3a, for the combinations of frequencies and doping densities (from $10^{10}$ to $10^{13}$ cm$^{-2}$) that allow relatively low loss propagation. As one can see the numbers are quite impressive, particularly in the mid-IR range of frequencies from 10 to 30THz where the nonlinear index can be as high as $10^{-4}$cm$^2$/W at lower doping densities. Of course, a significant part of this "giant" nonlinearity is simply due to division by the small (.33nm) thickness of the graphene monolayer. If one considers waveguide geometry, then the effective nonlinearity is somewhat less impressive –assuming $d_{eff} \sim \lambda/2$ one obtains $n_{2,eff} = 2N\alpha_0^2 v_F^2 / E_F \omega^3$ as shown in Fig.3b, assuming N=2 layers of graphene inside the guide. For operation around 10 μm (30THz) one would obtain $n_{2,eff} \sim 10^{-10} cm^2/W$ which is respectable, but not superior to what can be obtained in InAsSb using the scaling rules [28] for the same wavelength by operating near the half of bandgap energy. For the telecommunication band around 200 THz the effective nonlinearity in the waveguide geometry becomes about $n_{2,eff} \sim 10^{-13} cm^2/W$, i.e. comparable to the chalcogenide glass, although at this range, even below $2E_F$ the absorption can be quite high [26].

But let us turn our attention to the key parameter that characterizes nonlinearity – what is the optical length (number of layers in case of graphene) required to achieve the $\pi$ phase shift for switching. Clearly, the maximum change of refractive index that can be achieved is roughly $n_{eff,\omega}/2$ at input intensity of $I_{sat,\omega}$ - beyond that the change of index essentially saturates, just as in any nonlinear material [27]. The phase shift achieved in $N_l$ layers of graphene is then $\Delta\Phi = n_{eff,\omega} N_l d_{eff} (\omega/2c) = \alpha_0 N_l (E_F/\hbar\omega)$ and the number of layers necessary for the π-shift is $N_\pi = \pi \alpha_0^{-1} (\hbar\omega/E_F)$, shown in Fig. 4.a. Also shown in Fig.4.b is the switching power which we evaluate as a saturation intensity contained in a diffraction spot-size $P_{sw,\omega} = \pi\lambda^2 I_{sat,\omega}/4 = 2\pi^3 (E_F/e)^2 (c/v_F)^2/\eta_0$ which is actually frequency-independent, and depends solely on the density of electrons $P_{sw,\omega} = \frac{1}{2}\pi^3 \alpha_0^{-1} \hbar c^2 N_{2D} \approx 2\times 10^{-10} N_{2D} (W)$. The results are of course discouraging. For instance to achieve switching at $\lambda = 5\mu m$ one would need a carrier density of at least $10^{11} cm^{-2}$ and still require over 1000 layers for switching with a switching power in excess of 20W. Of course, doping 1000 layers of graphene to $10^{11} cm^{-2}$ using a single gate would require a field of $10^8$V/cm and probably cause a breakdown. The situation is even more bleak when one approaches telecommunication wavelengths for which both doping density and switching power would have to be increased, the latter to about 1kW.

Alternatively, one may consider the waveguide geometry. If we assume that the effective thickness of the waveguide is on the order of $d_{eff} \sim \lambda/2$ and the transparency condition $\hbar\omega \leq 2E_F$ is maintained the switching length required to achieve π-shift is $L_\pi \geq \pi\alpha_0^{-1}\lambda/N$ or longer than 200 free space wavelengths for two layers of graphene as shown in Fig.4c The switching power, also shown in Fig.4c will be $P_{sw,\omega} = \frac{\pi^2}{8}\alpha_0^{-1}\hbar\omega^2(c/v_F)^2 \approx 0.06W \cdot f^2$ where frequency $f$ is in THz. For 10 μm radiation a switching power of 50W would be required and for the telecom range it would reach 2.5kW – attainable only with low duty cycle mode-locked lasers and making all-optical signal processing impractical. It is also well known that to achieve high efficiency of other third order nonlinear effects such as frequency conversion via cross-phase modulation or four wave mixing, continuum generation, and others, one needs to satisfy essentially the same condition as optical switching, i.e. $n_{2\omega}I_\omega L/\lambda \sim \pi$ hence the same numbers of layers, lengths and optical powers as shown in Fig.4 are required to achieve a decent efficiency of all third order nonlinear phenomena in graphene.

Note the entirely different reasons for why graphene, despite nominally high nonlinear refractive index is not capable of being a medium for all-optical switching for the band-to-band and intraband (free carrier) processes. For the band-to-band process in undoped graphene (Fig1.a) the limiting factor is absorption and in terms of the figure of merit, the ratio of nonlinear index to linear absorption graphene is no different from any other semiconductor excited way above the bandgap. For free carrier transition, even if one discounts the free carrier and residual, phonon and defect assisted interband absorption, switching is hard to attain because only a relatively small number of carriers residing near the Dirac point exhibit nonlinear behavior, and even then the nonlinearity saturates once the carriers start oscillating with large amplitudes. **Essentially, all the electrons residing further away from the Dirac point hardly contribute to nonlinearity at all, but they are needed to keep the Fermi level high enough to mitigate** **the loss due to band-to-band absorption**. Clearly, absence of a real bandgap (rather than one induced by blocking the absorption) is a handicap that seems to have no remedy. It is also important to mention here that we have considered only the low temperature case. It is easy to see that at higher temperatures nonlinearity is reduced as carriers are promoted farther away from the Dirac point while the residual absorption increases due to phonon-assisted processes.

In the end, the nonlinear optical properties of any material depend on a very limited set of parameters – transition dipole matrix elements, density of states, and scattering (dephasing) rates, and none of these parameters is strikingly different in graphene from other materials, hence, in a hindsight, one should not be surprised that when it comes to practical nonlinear optical devices graphene exhibits no particular advantage over more conventional materials, such as SiN, Si, or chalcogenides. That does not preclude graphene from being used in a few specific niches, such as in a saturable absorber for mode-locked lasers, but that is due mostly to its easy availability for a particular wavelength range rather than to superior nonlinear figure of merit.

In conclusion we have developed an analytical expression for the nonlinear index of graphene and using it have estimated the doping densities, optical powers and number of layers (normal geometry) or interaction length (waveguide geometry) required to achieve a high efficiency of third order nonlinear effects at different wavelengths. The results do not show any advantages held by graphene over other nonlinear optical materials when it comes to the practical figures of merit, but one also cannot say that graphene is dramatically inferior to the other materials. Whether or not graphene becomes a material of choice will probably depend on considerations of mechanical and thermal robustness, manufacturability and cost.

The Author acknowledges support of NSF DMR-1207245

**Figure Captions:**

**Figure 1** Band structure and **(a)** interband absorption and nonlinearity in the undoped graphene and InAs and **(b)** intraband (free carrier) nonlinearity in doped graphene and InAs.

**Figure 2 (a)** origin of free carrier nonlinearity in graphene. the color corresponds to the projection of electron velocity on x axis **(b)** Normalized conductivity function $F$ as function of $(k/k_F)$ and **(c)** as a function of $(k/k_F)^2$. Fitting curve $F_{\sigma 1}$ is shown as dashed line. **(d)** Surface current density ***J*** and **(e)** surface conductivity $\sigma$ as functions of the drift quasi-momentum $\mathbf{k_0}$ for different values of sheet density of carriers

**Figure 3** Effective nonlinear index as a function of frequency for different values of sheet density of carriers for **(a)** normal incidence and **(b)** waveguide geometries

**Figure 4 (a)** Number of graphene layers and **(b)** switching power required to achieve p phase shift in normal geometry for different values of $N_{2D}$. **(c)** Switching power and switching length for the waveguide geometry

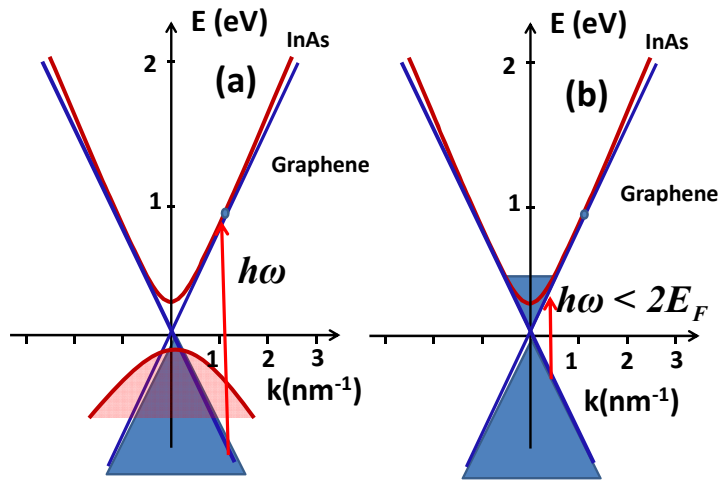

Fig.1

**Figure 1** Band structure and (**a**) interband absorption and nonlinearity in the undoped graphene and InAs and (**b**) intraband (free carrier) nonlinearity in doped graphene and InAs.

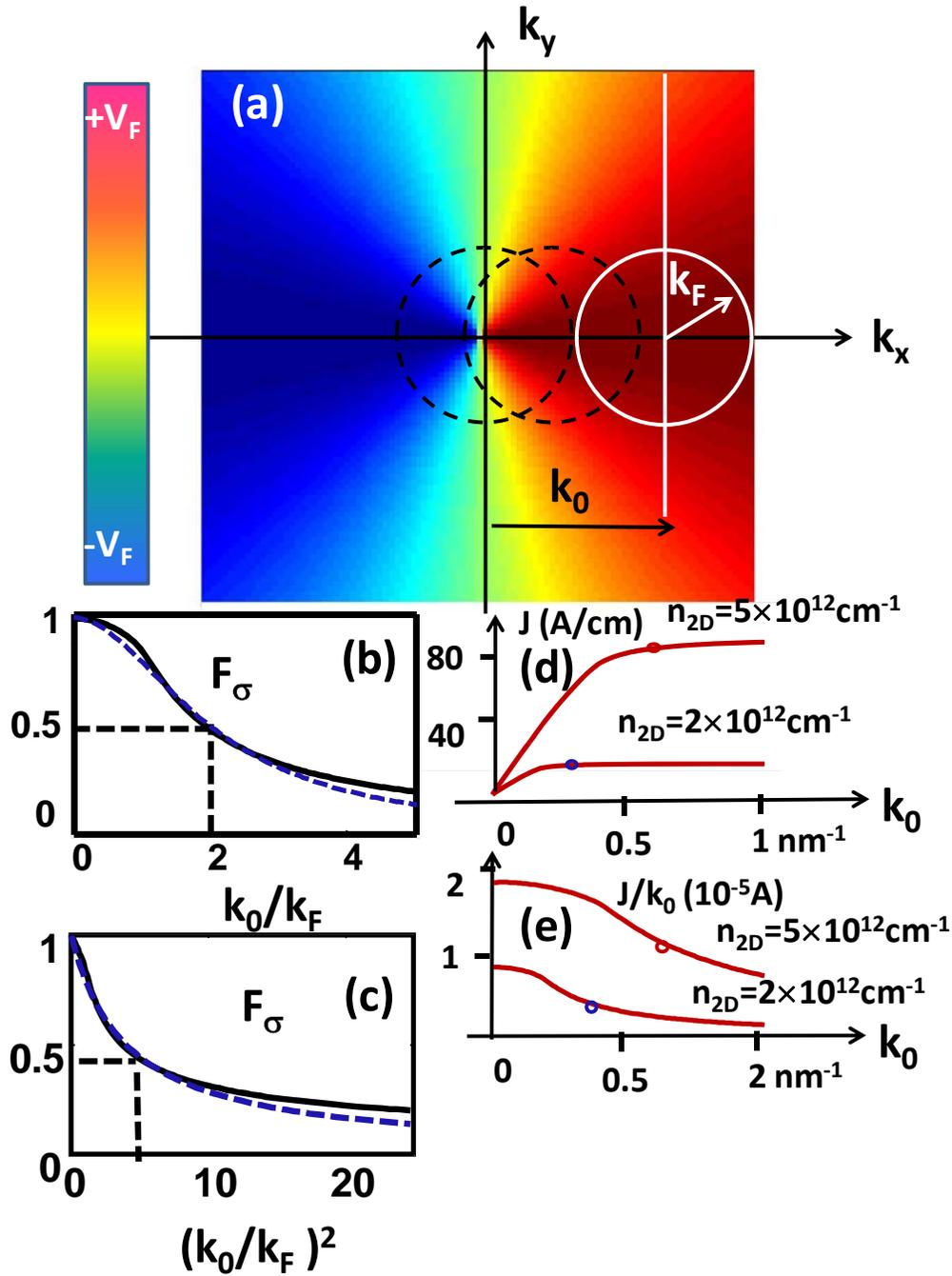

**Figure 2** (a) origin of free carrier nonlinearity in graphene the color corresponds to the projection of electron velocity on x axis (b) Normalized conductivity function $F_\sigma$ as function of $(k/k_F)$ and (c) as a function of $(k/k_F)^2$. Fitting curve $F_{\sigma 1}$ is shown as dashed line. (d) Surface current density $J$ and (e) surface conductivity $\sigma$ as functions of the drift quasi-momentum $k_0$ for different values of sheet density of carriers

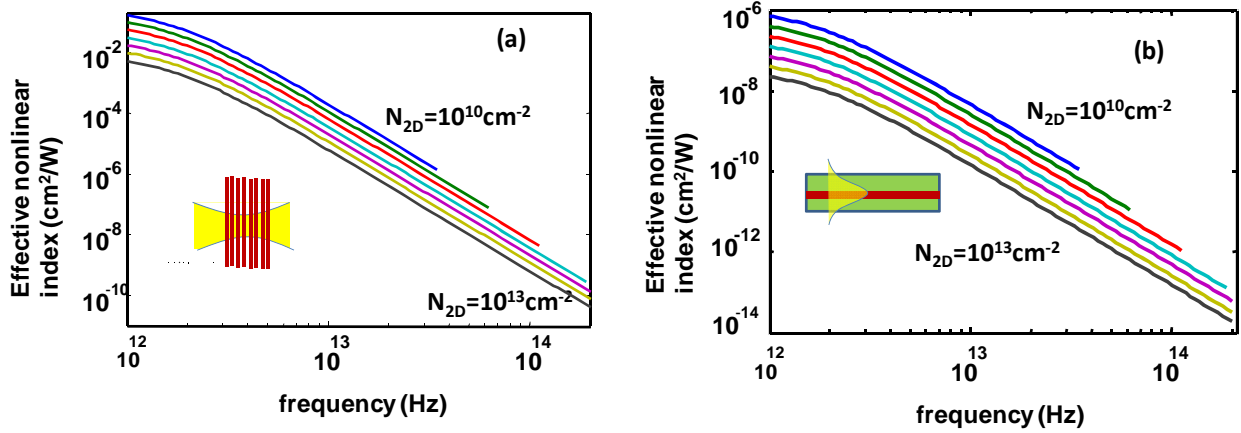

**Figure 3** Effective nonlinear index as a function of frequency for different values of sheet density of carriers for (a) normal incidence and (b) waveguide geometries

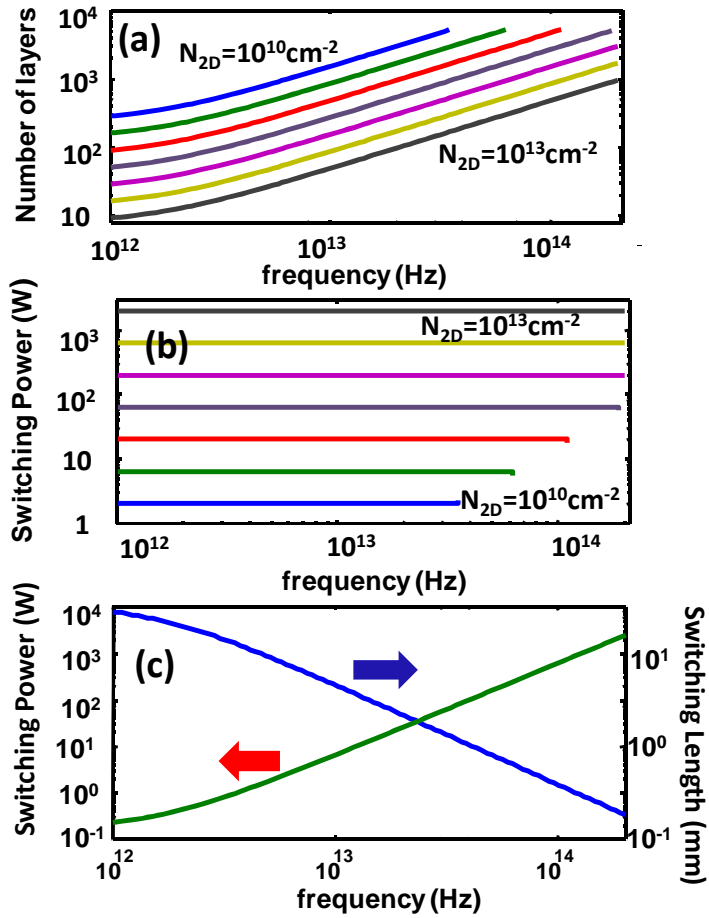

**Figure 4** (**a**) Number of graphene layers and (**b**) switching power required to achieve p phase shift in normal geometry for different values of $N_{2D}$. (**c**) Switching power and switching length for the waveguide geometry

Fig.4